\documentstyle[epsf,psfig]{mn}

\newcommand{\SS}{\scriptscriptstyle}
\newcommand{\Rl}{R_{\SS L1}}
\newcommand{\Md}{\dot{\cal M}}
\newcommand{\Msol}{{\cal M}_\odot}

\newcommand{\Rwd}{R_{wd}}
\newcommand{\SC}{\scriptsize}
\newcommand{\Teff}{T_{\mbox{\SC eff}}}
\newcommand{\TeffA}{T_{{\mbox{\SC eff}},{\rm A}}}

\newcommand{\cri}{{\mbox{\tiny crit}}}

\newcommand{\gdot}{^\circ_{\raisebox{.6ex}{\hspace{.05em}.}}}
\newcommand{\lsim}{{\textstyle{\; \lower 0.7ex\hbox{$<$}\;
  \atop \raise-0.1ex\hbox{$\sim$}}}}
\newcommand{\gcm}{g cm$^{-2}$}

\newcommand{\gs}{gs$^{-1}$}
\newcommand{\kms}{kms$^{-1}$}
\newcommand{\gsim}{{\textstyle{\; \lower 0.7ex\hbox{$>$}\; 
\atop \raise-0.1ex\hbox{$\sim$}}}}
\begin{document}

\title{Physical Parameter Eclipse Mapping of the quiescent disc in
V2051~Ophiuchi}

\author[Sonja~Vrielmann et al.]
{Sonja~Vrielmann$^1$\thanks{Send offprint requests to: S.\ Vrielmann},
Rae F.\ Stiening$^2$, Warren Offutt$^3$\\
$^1$Department of Astronomy, University of Cape Town, Private Bag,
Rondebosch, 7700, South Africa (sonja@pinguin.ast.uct.ac.za)\\
$^2$Department of Physics and Astronomy, University of Massachusetts,
Amherst, MA 01003-4525, USA (stiening@cannon.phast.umass.edu)\\
$^3$W\&B Observatory, P.O. Drawer 1130, Cloudcroft, NM 88317, USA
(offutt@apo.nmsu.edu)\\
}

\maketitle

\begin{abstract}
We analyse simultaneous UBVR quiescent light curves of the cataclysmic
variable V2051~Oph using the Physical Parameter Eclipse Mapping method
in order to map the gas temperature and surface density of the disc
for the first time. The disc appears optically thick in the central
regions and gradually becomes optically thin towards the disc edge or
shows a more and more dominating temperature inversion in the disc
chromosphere. The gas temperatures in the disc range from about
13\,500~K near the white dwarf to about 6\,000~K at the disc edge. The
intermediate part of the disc has temperatures of
9\,000~K to 6\,500~K.

The quiescent disc (chromosphere) shows a prominent bright region with
temperatures of 10\,500~K around the impact region of the stream from
the secondary with an extension towards smaller azimuths. The disc has
a size of $0.53 \pm 0.03 \Rl$ and a mass accretion rate of between
$\Md = 10^{15}$\gs to $10^{17}$\gs.  The light curves must include an
uneclipsed component, a hot chromosphere and/or a disc wind.

The PPEM method allows us to determine a new distance of
$146\pm20$~pc, compatible with previous rough estimates. For the white
dwarf we then reconstruct a temperature of 19\,600~K, if the lower
hemisphere of the white dwarf is occulted by the disc.

We suggest that the accretion disc is a sandwich of a cool,
optically thick central disc with hot chromospheric layers on both
sides as was suggested for HT~Cas. This chromosphere is the origin of
the emission lines.

We find that although V2051~Oph is very similar to the SU~UMa type dwarf
novae HT~Cas, OY~Car and Z~Cha, there must be a substantial difference
in order to explain its unique light curve. The reason for the
difference could either be a higher mass transfer rate caused by the
more massive secondary and/or a small but significant magnetic field of
the white dwarf, just strong enough to dissrupt the innermost disc.
\end{abstract}

\begin{keywords}
binaries: eclipsing -- novae, cataclysmic variables -- accretion,
accretion discs -- stars: V2051~Oph
\end{keywords}

\section{Introduction}

The eclipsing cataclysmic variable (CV) V2051~Oph was discovered by
Sanduleak (1972). Since then, the classification as a CV was quite
certain, but not, of which subtype. Bond \& Wagner (1977) suggested it
might be of AM Her type and this idea was supported by Warner \&
O'Donoghue (1987), hereafter WO, refining the classification to a
low-field polar. Other studies are based on the assumption that
V2051~Oph is a dwarf nova (Watts et al.\ 1986, Baptista et al.\ 1998)
with infrequent outbursts and low states (Warner \& Cropper 1983,
Baptista et al.\ 1998). V2051~Oph shows typical features of a high
inclination dwarf nova exhibiting an accretion disc, like double
peaked emission lines.

The mystery of V2051~Oph seemed to be solved with the detection of a
super outburst during which superhumps were detected (Kiyota \& Kato
1998). Since then, the system was classified as an SU~UMa type dwarf
nova. Therefore, we assume this system to exhibit an accretion disc.
This is further supported by Baptista et al.\ (1998) who do not find
typical features of a polar in their light curves of V2051~Oph, like
wanderings of ingress and egress phases or large amplitude pulsations.
However, in order to explain the complex behaviour of V2051~Oph,
Warner (1996) suggests that this system is a polaroid, an intermediate
polar with a synchronized primary.

With 1.5$^h$ it has a rather short orbital period, very similar to
HT~Cas, OY~Car and Z~Cha and in fact it is often compared to
these. However, the accretion disc seems to be highly variable, as
apparent from the variable eclipse profiles. Sometimes the disc seems
to be only small, hot and sharply defined, while at other times the
disc appears to be more extended (Warner \& Cropper 1983).
Furthermore, at times it shows a hump in the light curve with variable
timing (before or after eclipse) and variable occurance (it can occur
at one eclipse and disappear at the next) (WO). At other times, no such
hump is observed (e.g.\ this paper).

As V2051~Oph is an eclipsing system, the system parameters are quite
well known (e.g.\ Baptista et al.\ 1998). However, the lack of traces
of the secondary even in the infra-red prevents pinning down the
distance. The estimate of Watts et al.\ (1986) of a distance between 90 to 150 pc
is based on the assumption that the quiescent disc emits in an
optically thick fashion.  Berriman, Kenyon \& Bailey (1986) derive
values between 120 and 170~pc, allowing for optically thin material in
the disc.

One of our aims was to pin down the distance with the help of the {\em
Physical Parameter Eclipse Mapping} (PPEM, Vrielmann, Horne \& Hessman
1999, VHH) method as explained Vrielmann (2001) and briefly described
in Section~\ref{distance}.  The PPEM is a Maximum Entropy method (MEM,
Skilling \& Bryan 1984) designed to analyse accretion discs in CVs: by
analysing the shape of the eclipse light curve in various colours, we
can retrieve information about the physical properties within the
accretion disc. For this, we assume a certain spectral model for the
disc emission. The application of this method to quiescent data is
shown in Section~\ref{quiescence}. The last Section~\ref{discussion}
summarizes and discusses the results.

\section{The data}

The data set was taken by R.\ Stiening between 14 and 18 June, 1983
with the Mt. Lemon/USA 1.5m telescope. Attached to the telescope was
the SLAC (Stiening) photometer allowing the detection of light in the
four filters UBVR simultaneously. The data set was calibrated and
reduced by Keith Horne and consists of eight eclipses with a time
resolution of about 1 sec. A log of the data is
shown in Table~\ref{tab_data}. The procedure of the observations
is similar to that of HT~Cas by Horne, Wood \& Stiening (1991).

\begin{table}
\caption{The quiescent light curves of V2051~Oph.
\label{tab_data}}
\vspace{1ex}
\hspace{0.5cm}
\begin{tabular}{ccc}
\hspace{0.5cm} Date \hspace{0.5cm} & UT(start) & Eclipse No.\\ \hline
14/6/83 & 6:51:06 & 1,2\\
15/6/83 & 6:35:29 & 17\\
16/6/83 & 5:38:05 & 32,33\\
17/6/83 & 5:28:07 & 48\\
18/6/83 & 5:02:52 & 64,66\\
\hline
\end{tabular}
\end{table}

For the analysis of the quiescent data we assume a flat disc
geometry. This involves adjusting the out-of-eclipse levels to the
same level before and after eclipse by fitting a low order polynomial
to the out-of-eclipse data.

\section{The PPEM analysis}
\label{quiescence}

A preliminary analysis of the data set was shown in
Vrielmann~(1999). We arrive at some different results in the present
study because of redetermination of the ephemeris (using the 1983 data
alone and exclusion of two faulty R-band eclipses) and the correction of an
unfortunate error in the computer programme which changed the results
significantly. A slight difference to the result presented in
Vrielmann (2001) is due to a careful reexamination of the original data. This,
however, did not change the results significantly.

\subsection{The light curves}

\begin{figure*}
\hspace*{0.1cm}
\psfig{file=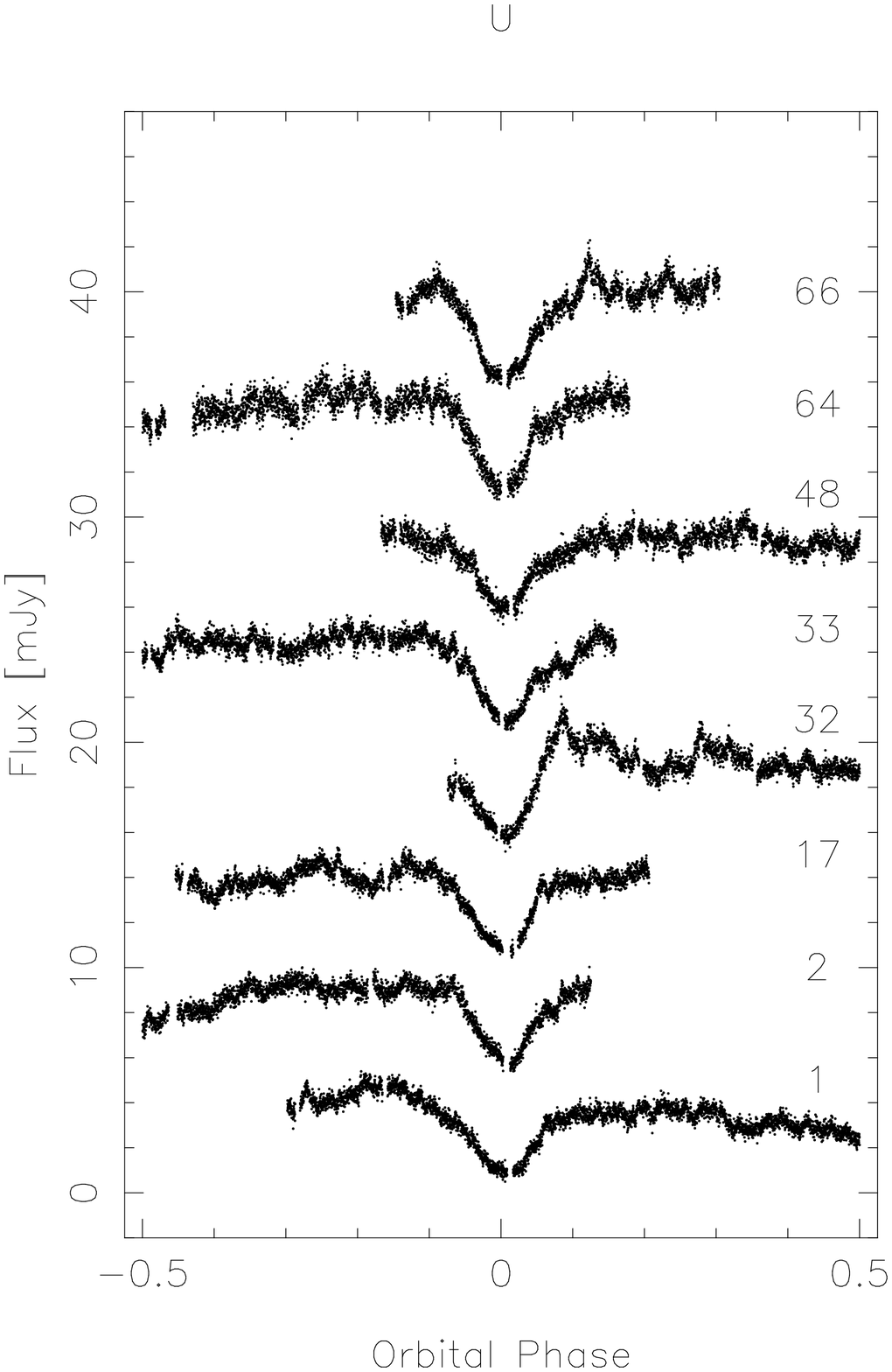,width=8cm}
\psfig{file=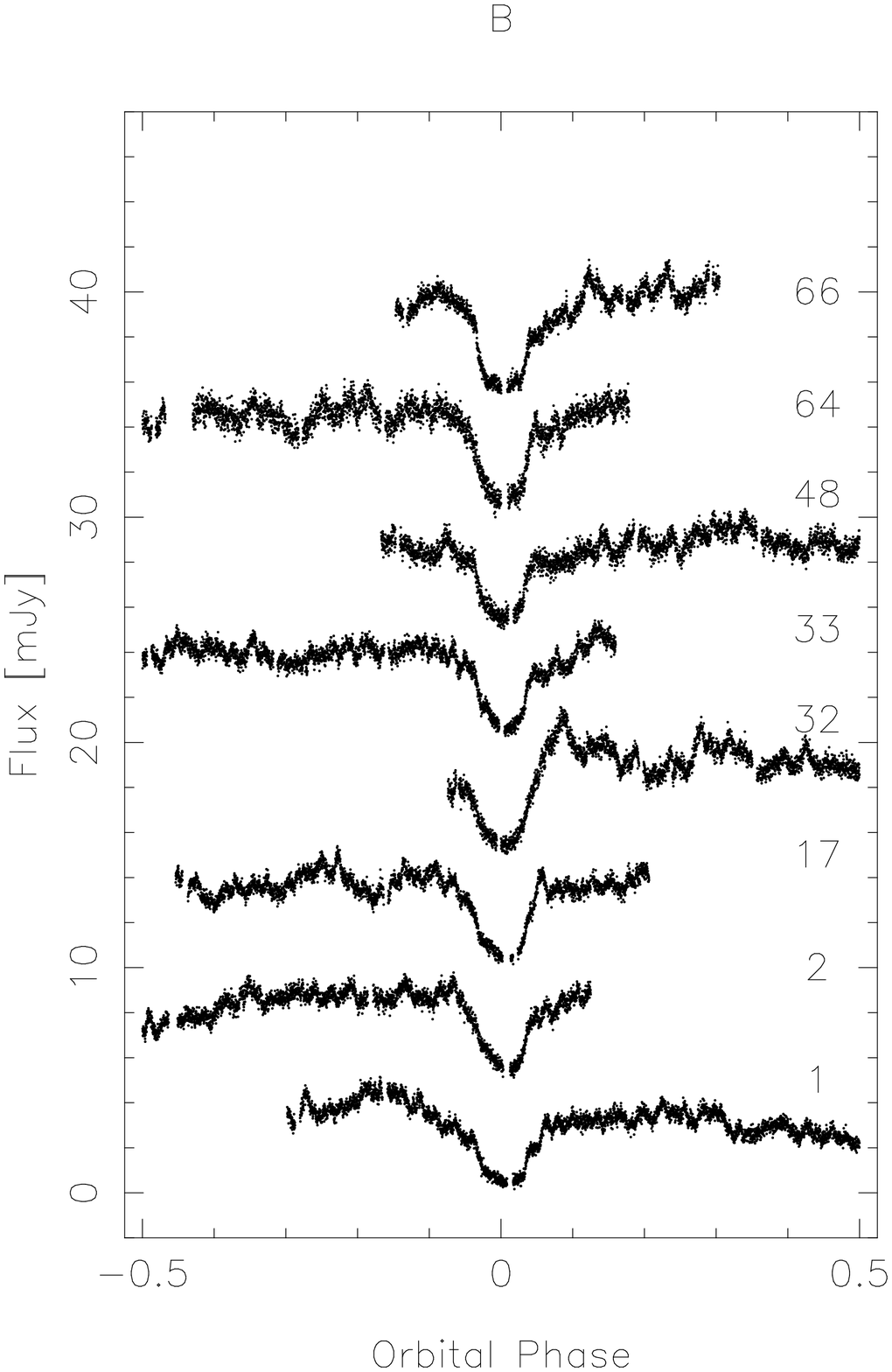,width=8cm}
\caption{\small The individual U-band and B-band light curves. The numbers on
the right give the eclipse numbers according to Table~\ref{tab_data}.
The eclipse light curves were offset by 5~mJy.
\label{oph_ub}}
\end{figure*}

\begin{figure*}
\hspace*{0.1cm}
\psfig{file=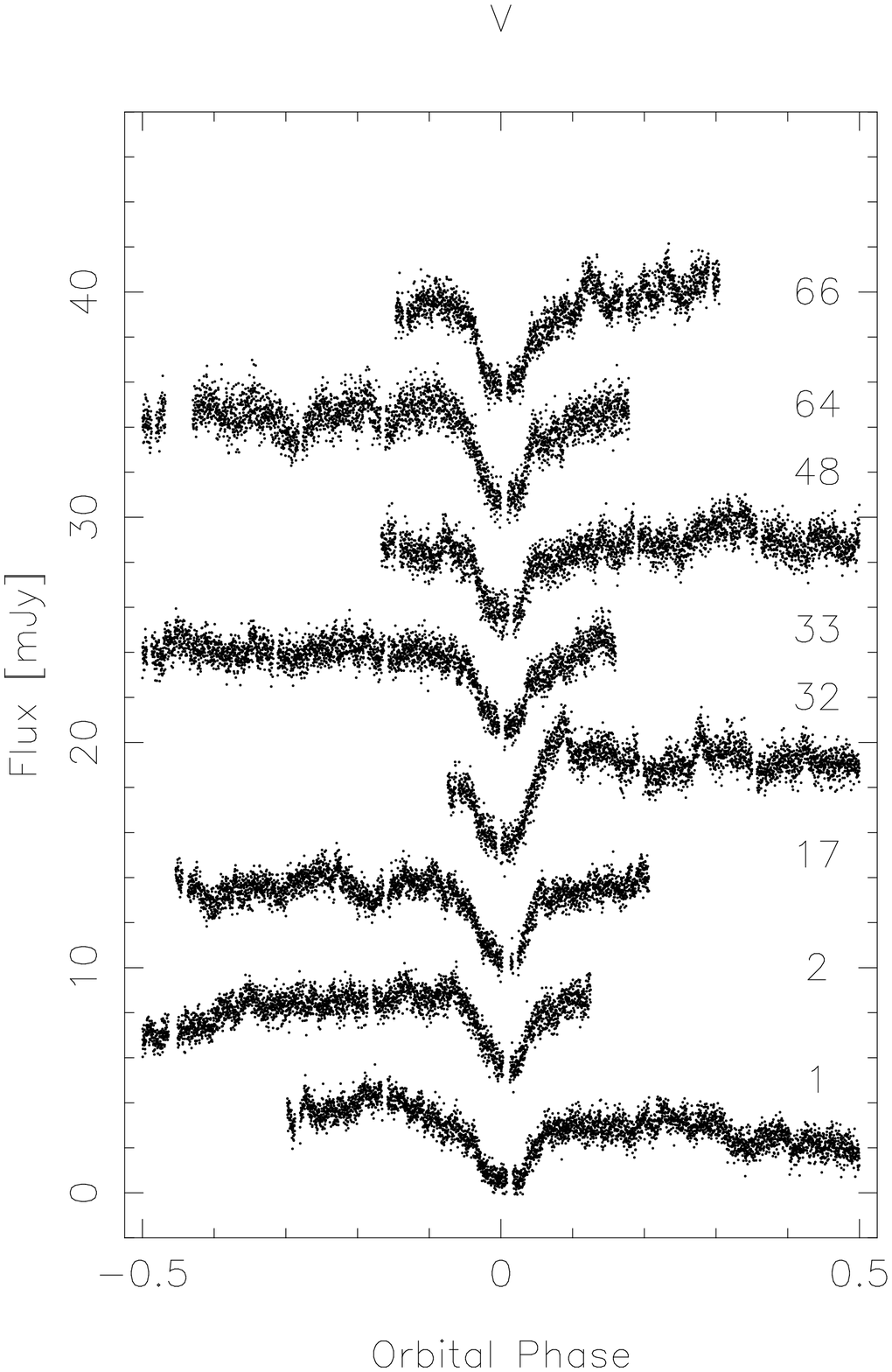,width=8cm}
\psfig{file=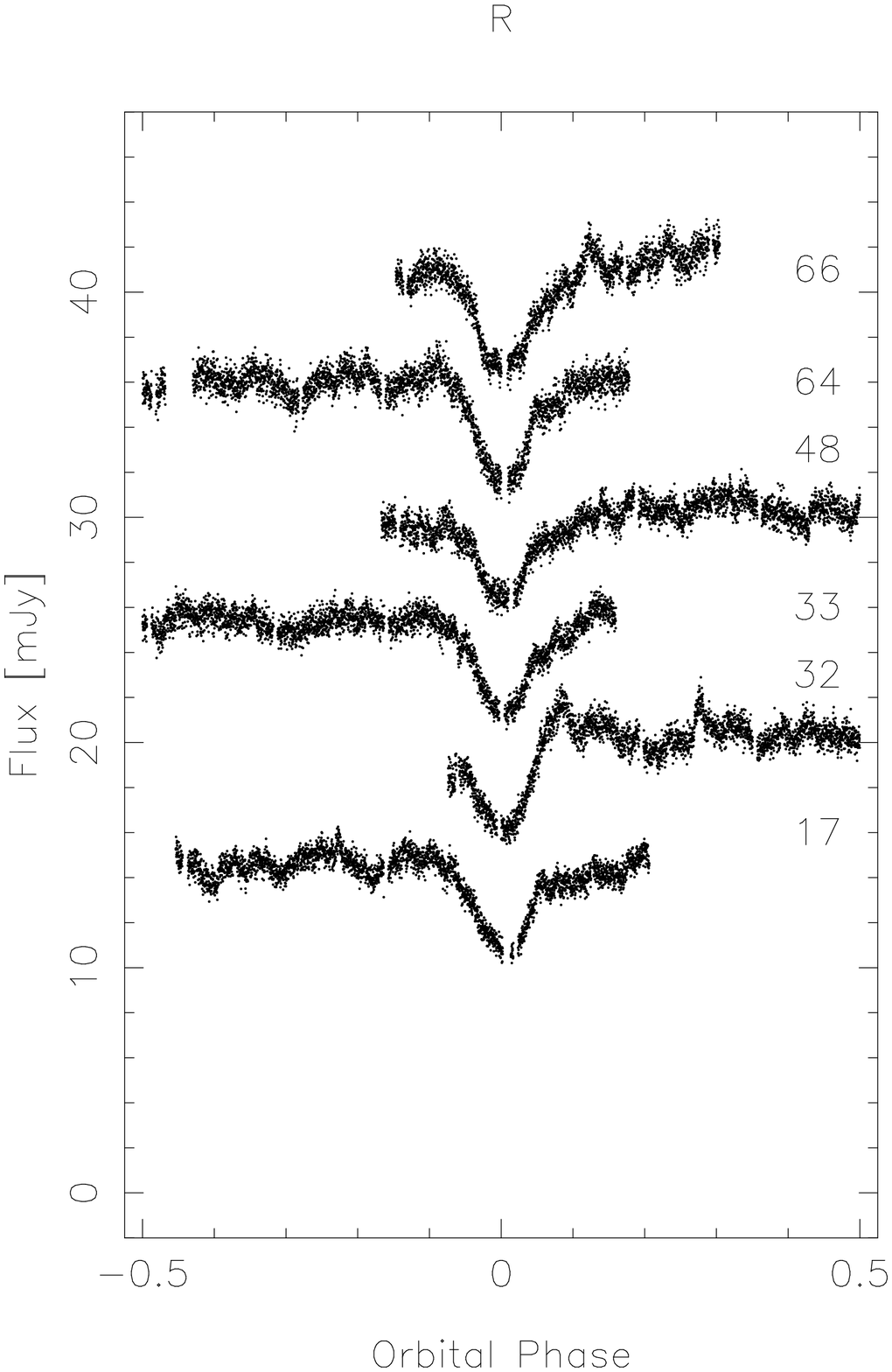,width=8cm}
\caption{\small The individual V-band and R-band light curves. Otherwise same as
Fig.~\ref{oph_ub}. The R-band light curve no.~1 and part of no.~2 is
excluded from the analysis due to missing sky measurements.
\label{oph_vr}}
\end{figure*}

Fig.~\ref{oph_ub} and \ref{oph_vr} show the individual UBVR light
curves of V2051~Oph at the full time resolution of 1s, showing the
eclipse of the white dwarf and accretion disc by the secondary. The
gaps are due to calibration measurements, unfortunately often during
mid-eclipse. However, since the ingress and egress contain most
information on the disc emissivity, this choice for calibration
measurements is a compromise one can live with. The R-band light curves
no.~1 and no.~2 were excluded from the analysis due to missing or faulty
sky measurements.

The phases were calculated according to a newly determined ephemeris
of
\begin{eqnarray}
HJD(eclipse) = 2445499.81007 + 0.062430039 \times E \\
BJD(eclipse) = 2445499.81002 + 0.062430039 \times E
\end{eqnarray}

using 26 out of the 30 light curves displayed in Fig.~\ref{oph_ub} and
~\ref{oph_vr} (eclipse no.~32 was excluded, because of the incomplete
coverage of the eclipse ingress). It was derived from the mean of the
steepest ingress and egress variation (at phases $\pm 0.033$) which
can be attributed to the eclipse of the white dwarf by the
secondary. Table~\ref{tab_omc} gives the (O-C) values with respect to
Baptista et al.'s (1998) ephemeris. Our (O-C) values are somewhat
large (although not unlike the scatter in other measurements) and not
in agreement with Baptista's (2001, private communication) cyclic
ephemeris. However, special care has been taken in deriving the
mid-eclipse times (e.g.\ including leap seconds) and the error in our
mid-eclipse times is certainly much smaller than the (O-C) values
given.

\begin{table}
\caption{The eclipse timings of our data. The cycle number and
observed minus calculated times are given with respect to Baptista et
al.'s ephemeris.
\label{tab_omc}}
\vspace{1ex}
\hspace{0.5cm}
\begin{tabular}{ccc}
Cycle & HJD & (O-C)\\
      & (2445000.+) & (cycles)\\ \hline
36103 & 499.81007 & -0.0123\\
36104 & 499.87249 & -0.0125\\
36119 & 500.80902 & -0.0107\\
36135 & 501.80779 & -0.0119\\
36150 & 502.74424 & -0.0114\\
36166 & 503.74321 & -0.0094\\
36168 & 503.86801 & -0.0103\\
\hline
\end{tabular}
\end{table}

\begin{table}
\caption{The white dwarf ingress and egress times and phases.
\label{tab_ine}}
\vspace{1ex}
\hspace{0.5cm}
\begin{tabular}{cccc}
HJD ingress & ingress phase & HJD egress & egress phase\\
(2445000.+) & & (2445000.+) & \\ \hline
 499.80800 & -0.033189 & 499.81214 & 0.033125 \\
 499.87044 & -0.033030 & 499.87454 & 0.032644 \\
 500.80696 & -0.031918 & 500.81108 & 0.034076 \\
 501.80574 & -0.033530 & 501.80984 & 0.032144 \\
 502.74221 & -0.033219 & 502.74628 & 0.031974 \\
 503.74106 & -0.033709 & 503.74536 & 0.035168 \\
 503.86594 & -0.033390 & 503.87007 & 0.032764 \\
\hline
           &  -0.03314 &           & 0.03313\\
\hline
\end{tabular}
\end{table}

While the white dwarf eclipse is relatively stable (see
Tab.~\ref{tab_ine}), the remainder of the light curve is quite
variable, showing flickering and flares which cause the accretion disc
ingress and egress profile to vary quite significantly. During white
dwarf eclipse the flares are less strong, though not completely
absent, indicating that the region where the flares originate is not
only confined to the immediate vicinity of the white dwarf, but also
extends somewhat into the inner disc.

The variation in the out-of-eclipse light curve is similar in all
filters, e.g.\ flares occur in light curves no.~32 (phase 0.09 and
0.28) or no.~66 (phase 0.12). In the V-band and R-band light curves the flares
appear slightly stronger, however, not as extreme as in HT~Cas (Horne
et al.\ 1991), where a strong flare present in the R-band appears
nearly absent in U-band.

It is interesting to note that even though we see strong variability
of the eclipse light curves in form of flickering and flares, we do
not see as strong and broad humps as in some of WO's light curves.
Even though our data show a lower S/N ratio such strong humps cannot
be hidden, but their absence reflect a seasonally changing behaviour of
V2051~Oph.  Since WO present their data in counts/sec without
comparison with a star of constant flux, it is difficult to compare
their data with ours.

WO also claim not to see the steep ingress and egress features of the
white dwarf, while we can certainly distinguish quite well the white
dwarf eclipse from the accretion disc eclipse in most individual light
curves. However, a close look at their eclipse light curves shows in
many cases a characteristic kink in the lower part of the steep
ingress profile, possibly the second contact phase of the white dwarf,
and in many cases a corresponding kink (third contact phase of the
white dwarf) in the egress profile.

In spite of the varying eclipse profile, we averaged the light curves
in order to reduce the flickering which would otherwise introduce
artefacts into the PPEM reconstructions. To enforce the fit of the
eclipse ingress, egress and the eclipse depth we chose a phase
resolution twice as high within the phase range $-0.1$ to 0.1 as
outside. The total phase range from $-0.15$ to 0.15 excludes some
flares, however not all, noticable especially at phases $-0.135$ and
0.13. The large variation of the out-of-eclipse light curve levels is
reflected in the error bars, which are therefore smaller during
eclipse than outside the eclipse profile.

\begin{figure}
\hspace*{0.1cm}
\psfig{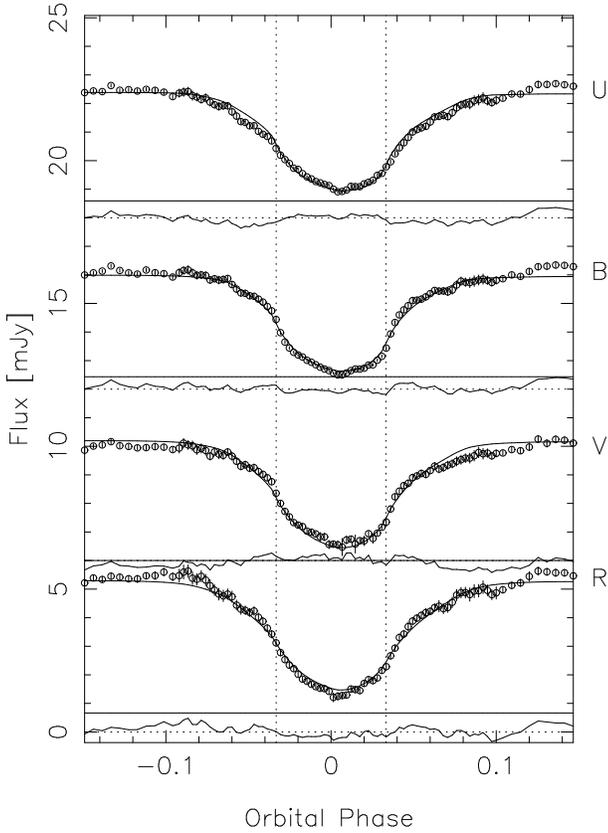}
\caption{\small The averaged quiescent light curves in UBVR together
with the fits. The ingress and egress phases of the white dwarf
(determined as steepest ingress and egress at phases $-0.033$ and
0.033) are drawn as vertical dashed lines. The data were offset by
6~mJy each with dotted horizontal lines indicating the zero-level for
each light curve and solid lines indicating the phase constant
uneclipsed flux (for the V-band light curve they coincide). Residuals are
displayed as solid lines in relation to the zero-level lines.
\label{ophf}}
\end{figure}

The averaged and phase-binned light curves are displayed in
Fig.~\ref{ophf}.  We fitted a low-order polynominal to the
out-of-eclipse light curves and divided the full light curves by the
polinomia in order to reduce any variation caused by an orbital hump,
though it was not very obvious in the averaged light curve.  An
orbital hump, as often seen in dwarf nova light curves, is usually
caused by anisotropic emission from the disc edge, in most cases
maximally seen just preceeding the eclipse. Since we used a
geometrically thin accretion disc geometry, we would not be able to
account for such an effect.

In comparison with the white dwarf ingress and egress profile, the
disc eclipse is very asymmetric. Within the white dwarf eclipse
(between phases $-0.033$ and $0.033$) the disc ingress is rather V
shaped, indicating a flat intensity profile, while the disc egress is
U shaped, indicative of a steep intensity profile. This is apparent in
all filters, though less pronounced in the noisier R-band light
curve. Furthermore, the flux at white dwarf ingress is higher than at
white dwarf egress. All these features are indicative of a brighter
leading lune of the disc, i.e.\ a bright spot where the ballistic
stream from the secondary hits the accretion disc.

\subsection{The PPEM method and the model}

The PPEM method is based on the eclipse mapping (EM) technique first
introduced by Horne (1985). In EM one takes advantage of the spatial
information of the intensity distribution of the accretion disc hidden
in the eclipse profile. By fitting the eclipse light curve of high
inclination systems, caused by the occultation of the accretion disc
by the secondary star, we reconstruct the intensity distribution using
a maximum entropy method (MEM). The MEM algorithm allows one to choose
the simplest solution still compatible with the data in this otherwise
ill-conditioned back projection problem. Further details of the EM
method can be found in Horne (1985) or Baptista \& Steiner (1993).

As described in VHH, the further developed PPEM method is designed to
reconstruct parameter distributions, like temperature $T$ and surface
density $\Sigma$, rather than intensity distributions of the accretion
disc. Hereby, the assumption of a spectral model is necessary,
relating the parameters to be mapped (e.g.\ $T$, $\Sigma$) to the
radiated intensity in a given filter $I_\nu = f(T, \Sigma)$. One has
to be cautious, however, as the parameters are not necessarily
everywhere well defined: only after the reliability of the
reconstructed parameters is established can the resulting maps be
interpreted.

We used the two-parameter spectral model described in VHH, to map the
temperature and surface density of the disc material. Though it is a
very simple, pure hydrogen model including only bound-free and
free-free H and H$^-$ emission, it is still very useful, because it
allows us to distinguish between regions of the disc that emit like a
black body and those which deviate from it. In the simplest
interpretation, the latter are optically thin regions and presumably
the site of line emission (Watts et al.\ 1986, Baptista et al.\
1998). However, real disc might show a Balmer Jump in absorption due
to decreasing temperature with disc height or a temperature inversion
in the disc photosphere (i.e. a disc chromosphere with line
emission). If such a more complicated scenario is present in the disc,
our simple solution will fail.

\begin{figure}
\hspace*{1cm}
\psfig{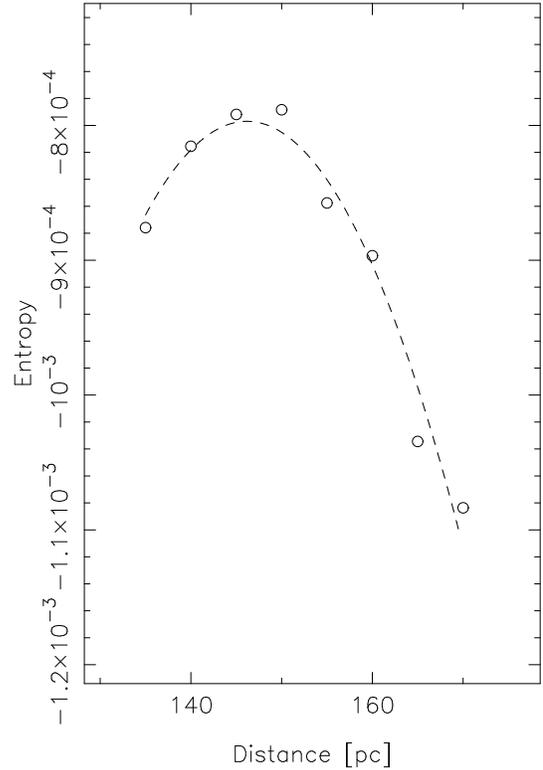}
\caption{\small The entropy of the reconstructions (as a measure of
smoothness) as a function of the assumed distance. The dashed line is
a parabolic fit to the data. It peaks at a distance of 146~pc.
\label{ophef}}
\end{figure}

For the white dwarf emission we used white dwarf spectra. As described
in VHH, we assumed the white dwarf to be spherical in terms of
occultation by the secondary and occultation of the accretion disc by
the white dwarf. Otherwise it is treated as a single pixel object and
we assign a single temperature to the white dwarf.
For the white dwarf we used pass band response functions of the UBVR
filters.

For reference of spatial structures in the disc, we use radius and
azimuth. The radius is used in units of the distance between the white
dwarf and the inner Lagrangian point $L_1$, i.e. the point $L_1$ has a
radius of $1\Rl$. The second coordinate is the azimuth, the angle as
seen from the white dwarf and counting from $-180^\circ$ to
$180^\circ$. Azimuth 0 points towards the secondary, the leading lune
of the disc has positive and the following lune negative azimuth
angles.

\subsection{The distance}
\label{distance}

\begin{figure*}
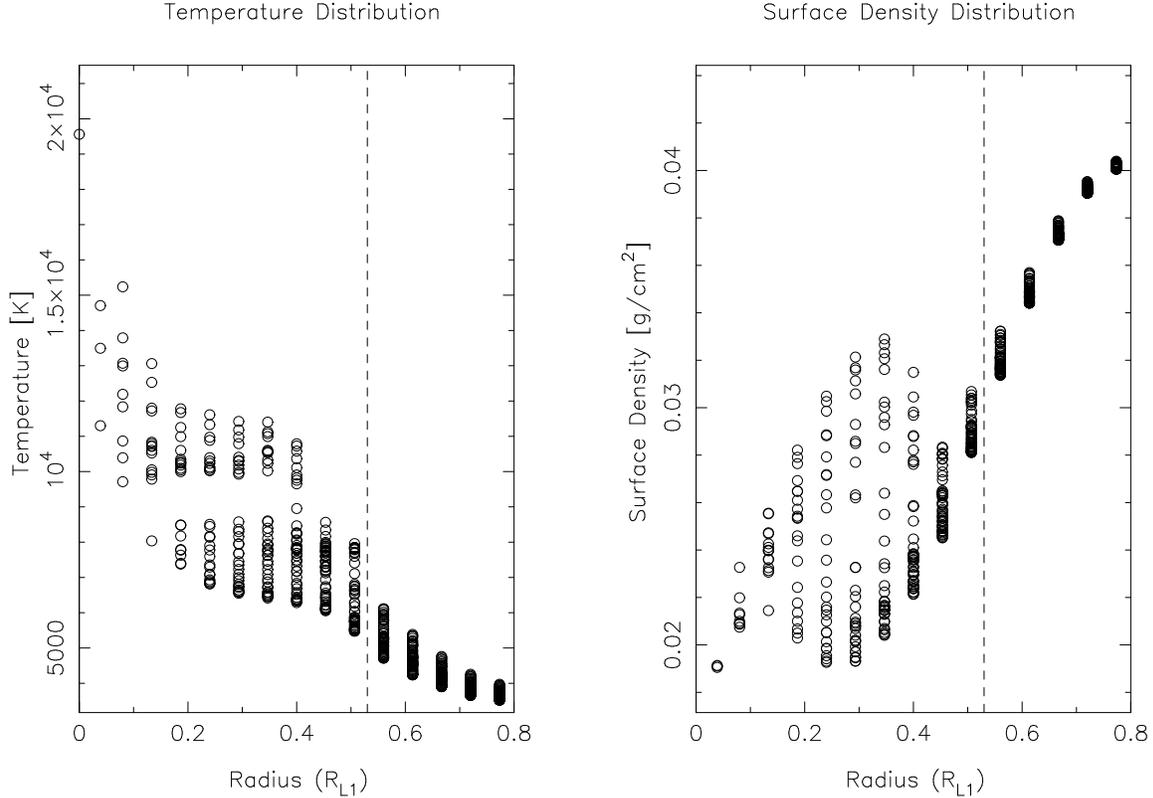

\hspace*{0.1cm}
\psfig{file=ophc145_25p1.ps,width=7cm}
\hspace*{1cm}
\psfig{file=ophc145_25p2.ps,width=7cm}
\caption{\small The reconstructed temperature ({\em left}) and surface
density ({\em right}) distributions.
\label{ophp}}
\end{figure*}

Before reconstructions of physical parameters can be made reliably,
the distance to the system must be known. Up to now, no reliable
distance could be determined, because the secondary was not detected,
not even in the infrared (Berriman et al.\ 1986). The distance
estimate of $d = 90$ to 150~pc exist with the assumption that the disc
is optically thick (Watts et al.\ 1986). Allowing for optically thin
emission in the disc, Berriman et al.\ find slightly larger distance
estimates of between 120 and 170~pc.  Line emission present in the
quiescent state (Baptista et al.\ 1998) suggests, that the disc is at
least partially optically thin (e.g.\ in the outer regions, Watts et
al.). A rough distance estimate of 184~pc, can be made using
the white dwarf flux (Catal\'an 1998, private communication).

Vrielmann~(2001) and Vrielmann, Hessman, Horne (2002) describe the procedure
how to arrive at a PPEM distance estimate and successful applications
of this method. Our method is based on the fact that at the best distance
the PPEM reconstructions are smoothest, expressed by a maximal
entropy. Fig.~\ref{ophef} shows the entropy as a function of the
distance which peaks at 146~pc. For our further study we use the trial
distance 145~pc which will give essentially identical results.

The exact value of the final $\chi^2$ of the fit to the observed data
depends on the data set used, e.g.\ the amount of flickering still
present after averaging or the applicability of the chosen spectral
model. No standard value can be given for this final $\chi^2$. It is
rather by choice of the analyst, judging the amount of structure in
the reconstructions, the performance of the mapping algorithm and the
quality of the fits and is therefore a rather subjective measure. It
is therefore usual not to fit to a $\chi^2$ of 1, but to a somewhat
higher value, allowing especially for deviations of the real spectral
characteristic from the chosen spectral model.

It is difficult to determine an exact error for the distance, because
of the uncertainty in difference between the spectral model used and
the true spectral characteristic. In a future study we will attempt to
quantify this error using more realistic spectra, such as calculated
by Hubeny (1991). For the time being we estimate the PPEM distance
error (with the assumption of a correct spectral model and good
representation of the observation) generously to be about 20~pc
according to some limited testing with small changes in the constant
part of the ephemeris (i.e. $\varphi_0$)\footnote{An error in
$\varphi_0$ shifts the white dwarf from the disc centre into the disc,
introduces artefacts in the reconstructed map and thereby influences
the entropy of the map as a function of trial distance.},
corresponding to about twice the best Hipparcos distance estimates for
objects with parallaxes less than 7~mas.

The inclination angle $i = 83.3^\circ$, the mass ratio $q = 0.19$, the
white dwarf mass ${\cal M}_{wd} = 0.78 \Msol$ and radius $R_{wd} =
0.0244 R_{\odot}$ as well as the scale parameter $\Rl/R_{\odot} = 0.42$
were taken from Baptista et al.\ (1998).

\subsection{The averaged light curves and final fits}

\begin{figure*}
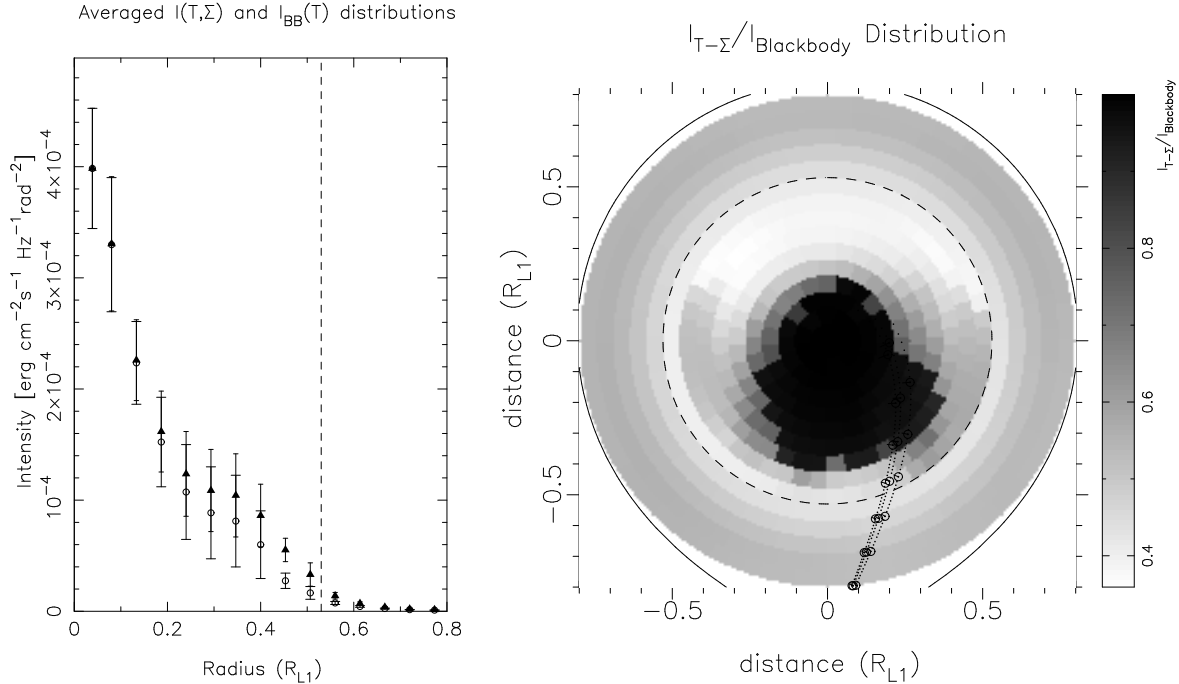

\hspace*{0.2cm}
\psfig{file=ophc145_25ib.ps,width=6cm}
\hspace*{0.3cm}
\psfig{file=ophc145_25iv.ps,width=9cm}
\caption{\small {\em Left:} The azimuthally averaged intensity
distributions $I(T,\Sigma)$ (open circles) and $I_{BB}(T)$ (filled
triangles) for comparison with error bars representing the scatter in
each annulus. The dashed line at 0.53 $\Rl$ indicates the size of the
disc. {\em Right:} The ratio $I_{T,\Sigma}/I_{BB}$ as a grey-scale
plot. The solid line represents the Roche-lobe, the dashed circular
line the size of the disc and the dotted lines ballistic stream
trajectories for mass ratios $0.19 \pm 50$\%. The secondary Roche lobe
is at the bottom, outside the plotted region. See text for further
explanation.
\label{ophbb}}
\end{figure*}

Fig.~\ref{ophf} shows the averaged quiescence light curves together
with the final fit ($\chi^2 = 2.5$) for a distance of 145~pc. While the
overall profiles are reproduced, the fits deviate in detail from the
observed light curves. This is mainly due to flares and flickering,
e.g.\ at phases $-0.135$ and 0.13. The U-band and B-band light curves are fitted
best, which means the Balmer jump is quite well reproduced. The
residuals of the R-band light curve are worst, most likely due to the lack
of line emission, particularly H$\alpha$, in our model spectrum.  This
agrees with the negative residuals at mid-eclipse and the too shallow
model light curve.

Further deviations, e.g.\ in disc ingress and egress, are probably due
to the simplicity of the spectral model used. Keeping in mind that all
light curves are fitted simultaneously, we would be rather surprised
that such a simple spectral model fits the data so well. A future
application of a more sophisticated, more realistic model would
probably give a better reproduction of the observed light curves.

\subsection{The reconstructions}

The reconstructed temperature and surface density distributions are
displayed in Fig.~\ref{ophp}. In the centre the temperature reaches
values up to nearly 15\,000~K. It is fairly constant in the
intermediate regions with a characteristic separation into a hot area
with $T\sim 10\,500$~K and a cooler region with $T\sim7500$~K. Towards
the disc edge it drops linearly. The surface density rises from the
central parts to the disc edge from values around 0.02 to 0.03~\gcm.

As pointed out in VHH, it is essential to study the behaviour of the
spectral model in the given parameter space, before one can interpret
the reconstructed maps. The easiest disc parameter to estimate is the
disc size. We just need to look at the intensity distribution derived
from the T-$\Sigma$ distributions, especially in the red filter. Since
the R-band light curve is fitted worst, however, we would rather consider the V-band
light curve in the present case. At a radius of 0.56$\Rl$ the V-band
intensity drops to about 1.5\% of the maximum (i.e.\ the white dwarf
intensity) and the distributions are to smoothed out due to the
maximum entropy constraint.  We therefore expect the disc edge to be
between $0.507 \Rl$ and $0.56 \Rl$, i.e. at $0.53 \pm 0.03 \Rl$. This
disc radius is compatible with being smaller than the 3:1 resonance
radius $R_{3:1} = 0.68\Rl$ (for $q=0.19$, Warner 1995), however,
significantly smaller than Steeghs et al.'s (2001) radius of $1.1\Rl$
(probably truncated due to tidal interactions to somewhere below
1$\Rl$). Our disc radius describes the disc emitting continuum
emission, while Steeghs et al.'s disc the line emission. It is
possible that the chromosphere of the accretion disc extends to larger
radii than 0.56$\Rl$, however, there is no indication (e.g.\
precession of the disc) that the quiescent disc is larger than 0.68$\Rl$.
We rather suggest that the disc as observed by Steeghs et al.\ was
affected by the immediately preceeding outburst.

When we compare the disc intensity $I(T,\Sigma)$ (averaged over
azimuth and wavelength) with the black body intensity derived from the
temperature distibution alone $I_{BB} (T)$ (Fig.~\ref{ophbb}), we see
in which parts the disc resemble black body emitters. In the inner part
of the disc the two distributions are nearly identical and the ratio
$I_{T,\Sigma}/I_{BB}$ is very close to unity, i.e.\ the disc is nearly
optically thick in the central region up to a radius of about
$0.15\Rl$. Towards the edge the disc become more and more optically
thin or the temperature inversion in the chromosphere becomes more and
more dominant.

The grey-scale plot in Fig.~\ref{ophbb} also shows that the nearly
optically thick region extends towards the region where the bright spot
is expected, i.e.\ where the accretion stream from the secondary hits
the accretion disc. In this impact region the disc material will have
a higher density compared with regions with the same radius but
different azimuth. However, the bright region extends also towards
smaller azimuths, i.e.\ spreads against the flow of matter in the
disc.

\begin{figure*}
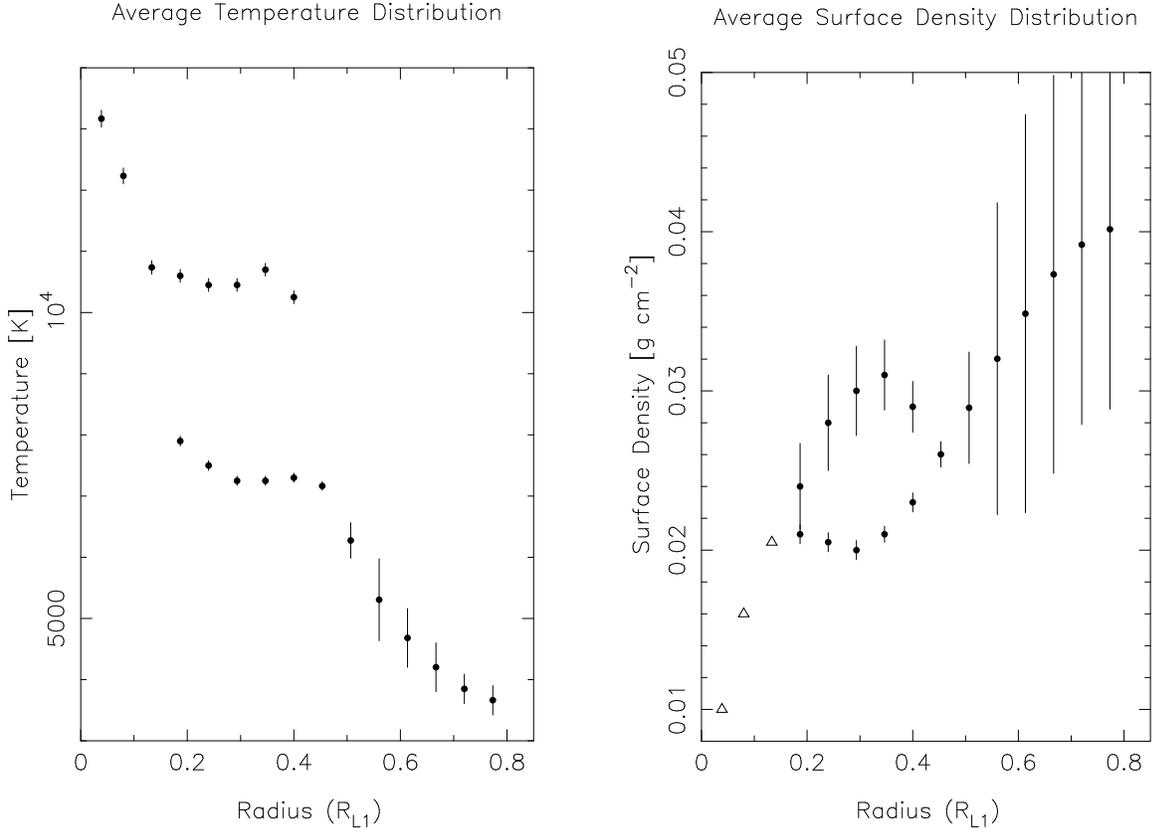

\hspace*{0.1cm}
\psfig{file=ophc145_25a1.ps,width=7cm}
\hspace*{1cm}
\psfig{file=ophc145_25a2.ps,width=7cm}

\caption{\small Averaged temperature ({\em left}) and surface density
({\em right}) distributions. The error bars indicate the reliability
of the reconstructed parameter values. At small radii the disc is
opticlly thick and we can only give lower limits for the surface
density, indicated by triangles.
\label{opha}}
\end{figure*}

For a more quantitative analysis of the derived parameters, we plot
the averaged parameter distributions in Fig.~\ref{opha} with error
bars according to a study of the spectral model. The error bars are
derived as follows: Similar to the study in VHH, we calculated the
spectrum for the averaged parameter pair ($T$,$\Sigma$) at a given
radius, calculated spectra for a number of parameter pairs ($T+\Delta
T$, $\Sigma + \Delta \Sigma$) around that parameter pair and
determined the deviation of the spectra in terms of the $\chi^2$. The
error bars include the temperature and surface density range for a
$\chi^2$ of 3 for the given parameter pair. Small error bars indicate
a high sensitivity of the spectrum to the parameter, while large error
bars mean only little variation in the spectrum in the given parameter
range, e.g.\ in the optically thick range the surface density would
have an infinitely large error bar.

The error bars given in the averaged distributions (Fig.~\ref{opha})
are good estimates for the error bars that should be attached to the
values in the original distribution (Fig.~\ref{ophp}). They are
omitted in Fig.~\ref{ophp} mainly for clarity of the plot, but partly
also because it would suggest an exact estimate of the error bars.

We split the intermediate part of the disc (radii 0.2 to 0.5$\Rl$)
into an ``upper'' and a ``lower'' branch. The upper branch corresponds
to the bright spot region, where the disc is nearly optically thick,
the lower to the other parts of the disc.

The temperature is within the disc radius very well constrained.  In
contrast, the surface density values are partly much less well
defined. The bright spot region has large error bars, because it is
nearly optically thick. At small radii the disc is optically
thick and therefore we can only give lower limits for the surface
density $\Sigma$. However, the true values of $\Sigma$ can be very
different from the given lower limits.

The temperature drops with radius as expected. It reaches about
13\,000~K in optically thick, inner part of the disc and drops to about
6\,000~K at the disc edge. The optically thick  bright region has
temperatures around 10\,500~K while the other regions with identical
radius have temperatures between 9\,000~K and 6\,500~K. These
temperatures are significantly lower than those quoted by Berriman et
al.\ (1986) of 20\,000~K.

The surface density most likely increases with radius (as suggested by
Fig.~\ref{ophp}) or could be roughly constant, considering the partly
large error bars in Fig.~\ref{opha}. (However, there is a clear
distinction between the bright spot region and the rest of the disc.)
Such a $\Sigma(r)$ distribution is compatible with theoretical
calculations by Ludwig, Meyer-Hofmeister \& Ritter (1994) and
Cannizzo, Gosh \& Wheeler (1982), but contradictory to Meyer \&
Meyer-Hofmeister's (1982) calculations. Since the critical surface
density distributions within the disc instability model show an
increase of $\Sigma_\cri$ (e.g.\ Cannizzo \& Wheeler 1984) we would
rather expect that the actual $\Sigma(r)$ distribution also increases
with radius. However, as we see later (Section~\ref{vertical}), these
surface densities probably describe only the chromosphere of the
disc.

\subsection{The white dwarf}

Using white dwarf spectra, the white dwarf temperature was
reconstructed to 19\,600~K using log g=8 white dwarf spectra. For the
geometry we used only the ``upper'' half of the white dwarf, i.e.\
$0.25 + (90^\circ - 83.3^\circ)/360^\circ = 0.269$ of the total
spherical surface. In the case that the disc is disrupted in the
central part (i.e.\ if V2051~Oph is an intermediate polar) we would
see more of the whole white dwarf and reconstruct a lower temperature. In
case the white dwarf were fully visible the temperature would drop to
15\,000~K. On the other hand, an accretion curtain may also hide part
of the white dwarf surface, making it difficult to determine an exact
value for the white dwarf temperature.

We estimated the error of the white dwarf temperature $T_{wd}$ by
checking how the light curve fit changes with different values of
$T_{wd}$ if we keep the disc parameters as reconstructed. This
gives us an error of about 1\,000~K. However, the error in $T_{wd}$ is
linked to the error in the distance. Taking this into account, the
total error is rather 2\,000~K.

The geometry we used does not specifically set a location or size of a
boundary layer. However, if such a boundary layer is present and
emitting in the observed passbands, the emission should either be
reconstructed in the innermost ring around the white dwarf or be
hidden in the white dwarf emission. That we do not see a specific
boundary layer emission in the innermost ring of the disc either means
that the white dwarf temperature is overestimated or that there
is no boundary layer (see Section~\ref{typesystem}).

Steeghs et al.~(2001) and Catalan et al.~(1998) both find a white
dwarf temperature of 15\,000~K. However, since both do not state which
distance value they used a direct comparison to our value is not
possible. Assuming they used a similar distance, this would mean that
we might see part of the ``lower hemisphere''. However, this is
doubtfull, since the disc is optically thick near the white dwarf and
should occult the ``lower'' half of the compact source. If the
boundary layer were hidden in the white dwarf emission, Steeghs et
al.\ and Catalan et al.\ should have faced the same problem.

The reconstructed value is close to typical values of white dwarf
temperatures, but somewhat higher than in the similar objects HT~Cas,
OY~Car and Z~Cha of around 15\,000~K (G\"ansicke \& Koester 1999 and
references therein). This might be connected to the somewhat different
system parameters as shown in Tab.~\ref{tab_para}.

\subsection{The uneclipsed component}
\label{sec_unecl}

\begin{table}
\caption{The reconstructed fluxes (in mJy) for the uneclipsed
component $F_{un}$ as well as a M4.5 and M5.5 main sequence star,
$F_{\mbox{M4.5}}$ and $F_{\mbox{M5.5}}$, respectively.
\label{tab_ucq}}
\vspace{1ex}
\hspace{0.5cm}
\begin{tabular}{cccc}
\hspace*{0.1cm} Filter \hspace{0.1cm} & \hspace{0.1cm} $F_{un}$
\hspace{0.1cm} & \hspace{0.1cm} $F_{\mbox{M4.5}}$ \hspace{0.1cm} &
\hspace{0.1cm} $F_{\mbox{M5.5}}$ \hspace*{0.1cm}\\ \hline
 U          & 0.59 &      &      \\
 B          & 0.44 &      &      \\
 V          & 0.00 & 0.09 & 0.01 \\
 R          & 0.66 & 0.30 & 0.07 \\
\hline
\end{tabular}
\end{table}

Apart from the disc parameters we also reconstructed an uneclipsed
fraction of the light curve (Table~\ref{tab_ucq}). For the
reconstruction of this component, we do not need to assume any
spectral model. It is simply any uneclipsed flux in the given filter
and free for any interpretation.

Since the uneclipsed spectrum might be generaly underestimated, the
B-flux overestimated (Vrielmann \& Baptista 2002) and since the errors
in the uneclipsed component particularly in the V and R-band filters
are rather large (Vrielmann et al.\ 2002), we cannot derive much
physics out of this component. All we can say is that the uneclipsed
component is compatible with a M5 main sequence star (using Baptista
et al.'s mass and radius and Kirkpatrick \& McCarthy 1994 tables for
main sequence stars, see Table~\ref{tab_ucq}) plus an optically thin
component emitting particularly in the U-band and R-band, e.g. a
chromosphere and/or disc wind.

Evidence for a disc wind comes from the fact that the blueshifted
Doppler emission component of all emission lines is always brighter
than the redshifted component (Kaitchuck et al. 1994, Honeycutt,
Kaitchuck \& Schlegel 1987). Even though Steeghs et al.~(2001) claim
to find a reversal of this asymmetry for H$\alpha$ (i.e.\ the red peak
is stronger) we cannot see this in their gray-scale plot: their Fig.~1
shows the blue wings stronger at all plotted phases (-0.3 to 0.1)
except during part of the eclipse. Furthermore, the eclipse is shorter
and shallower in the blue peak than in the red peak, indicating that
the origin of the blue emission is spatially focused and only partly
eclipsed, e.g.\ like a collimated wind (or jet) above the disc. And
although in Warner \& O'Donoghue's (1987) spectrum the red peak is
higher, the blue peak is slightly broader, indicating that the
additional emission in the blue wings has rather high velocities.

\subsection{Derived parameters}
\label{derived}

From the reconstructed parameter distributions we can calculate
several further parameters, like the effective temperature, the disc
scale height and the viscosity. These parameters are more descriptive
of the disc than the gas temperature and surface density.  The disc
size and effective temperature can in particular be determined more
accurately than the originally reconstructed ones, because they rather
depend on the intensity distributions corresponding to the original
reconstructions than on the reconstructed parameters themselves.
However, in general the derived values are only as certain as the
originally reconstructed values on which they are based and one has to
be careful in interpreting the resulting derived parameter values. One
must also keep in mind, that these parameters might only describe the
hot chromosphere, not the entire disc (see Section~\ref{vertical}).
Furthermore, it is difficult estimating an error of the derived
parameters without extremely time consuming Monte-Carlo type
calculations.

\subsubsection{The effective temperature and mass accreion rate}
\label{sec_teff}

\begin{figure*}
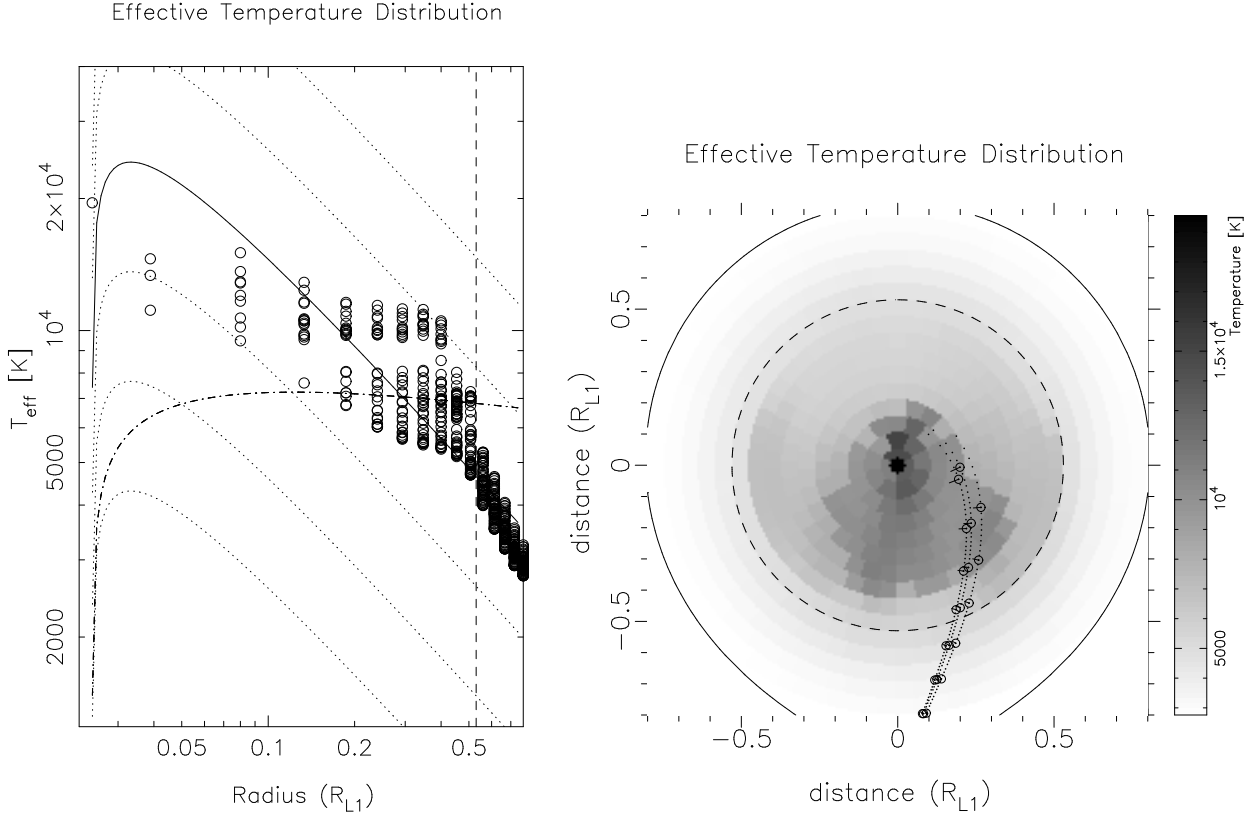

\hspace*{0.1cm}
\psfig{file=ophc145_25t.ps,width=7cm}
\hspace*{0.3cm}
\psfig{file=ophc145_25tv.ps,width=9cm}
\caption{\small {\em Left:} The radial effective temperature
distribution in quiescence. The underlying dotted lines are
theoretical steady state temperature distributions for $\log \Md = 13$
to 18, the one for $\Md = 10^{16}$\gs is drawn solid for reference.
The dashed-dotted lines indicates the critical effective temperature
\protect{$\TeffA$} according to Ludwig, Meyer-Hofmeister \& Ritter's
(1994) critical mass accretion rate $\Md_{\rm A}$. {\em Right:} The spatial
distribution of the effective temperature as a grey-scale plot. The
solid line represents the Roche-lobe of the primary, the dashed line
(in both plots) the disc size and the dotted lines ballistic accretion
stream lines for mass ratios 0.19 $\pm$ 50\%. The secondary Roche lobe
is at the bottom, outside the plotted region.
\label{ophteff}}
\end{figure*}

From the temperature and surface density distributions we can
calculate the spatially resolved effective temperature $\Teff$
distribution in the disc (Fig.~\ref{ophteff}). $\Teff$ is calculated
from the intensity $I_\nu(T,\Sigma)$ as
\begin{equation}
\pi \int I_\nu(T,\Sigma) \,\,d\nu = \sigma \Teff^4
\label{eq_teff}
\end{equation}
with $\sigma$ the Stefan-Boltzmann constant.  In practice, $I_\nu$ is
summed numerically over a large frequency range $\nu$.
Equation~\ref{eq_teff} shows that $\Teff$ does not directly depend on
the reconstructed values $T$ and $\Sigma$, but rather on the intensity
derived from them. Since the intensity distribution directly translates
into the observed fluxes which are as good as the fit to the light
curves suggests, the effective temperature is more reliable than the
surface density values and rather comparable to the certainty of the
reconstructed temperature values.

The main part of the disc has a mass accretion rate of between
$\Md = 1\times 10^{15}$~\gs\ to $\Md = 1\times 10^{17}$~\gs\
($1.6\times10^{-11}\Msol$~yr$^{-1}$ to
$1.6\times10^{-9}\Msol$~yr$^{-1}$). However, the distribution is split
into two parts of higher and lower temperatures. As the grey-scale
plot in Fig.~\ref{ophteff} shows, the higher temperatures are
associated with the bright spot region, or more general to the region
between the two stars, i.e.\ the same region that is nearly optically
thick. The remainder of the disc is cooler.

This situation is understandable in terms of the mass accretion rate
involved. At the bright spot the mass transfer is highest, reaching
$1\times 10^{17}$~\gs, because of the impact of the gas stream from
the secondary. In the remaining disc, the mass accretion rate is
determined solely by the viscosity. In the region $180^\circ$ away
from the gas stream the mass accretion rate drops to values around
$3\times 10^{15}$~\gs.

Fig.~\ref{ophteff} also shows the theoretical critical effective
temperature $\TeffA(R)$ derived from the critical mass accretion rate
$\Md_{\rm A}(R)$ as given by Ludwig et al.\ (1994), indicating whether
the disc should undergo outbursts (see Section~\ref{massacc}).

\subsubsection{The disc mass}

The disc mass can be estimated from the surface density, however, only
with a certainty that the reconstructed surface density values
allow. For the disc within a radius of 0.53$\Rl$ we derive a mass
of $7.7 \times 10^{19}$~g or $3.9\times10^{-14} \Msol$.
This value has to be treated with some caution. The disc is nearly
optically thick in the centre, i.e.\ for that region we rather derive
a minimum value of $\Sigma$ and therefore of the disc mass. In the
outer parts of the disc the surface density might vary by up to a
third of the reconstructed value for $\Sigma$. The error in the disc
mass therefore must be set to roughly 50\% of the derived value.

According to Frank, King \& Raine (1992) the mass in the disc at any
one time is $M_{disc} \lsim 10^{-10}\Msol \alpha^{-4/5}
\Md^{7/10}_{16}$, where $\Md_{16}$ is the mass accretion rate in units
of $10^{16}$ \gs\ and $\alpha$ the viscosity coefficient determined by
$\nu = \alpha c_s H$, where $\nu$ is the viscosity, $c_s$ the sound
speed and $H$ disc thickness. So, unless $\alpha$ reaches unreasonably
high values of the order of $10^4$ or the underlying disc is extremely
massive, Frank et al.'s condition is well fulfilled.

\subsubsection{The disc scale height}

\begin{figure}
\hspace*{2cm}
\psfig{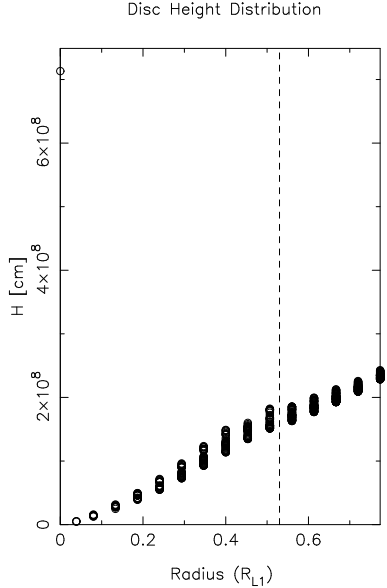}
\caption{\small The radial distribution of the disc scale height. The
circle at a disc radius of 0$\Rl$ represents the radius of the white
dwarf and illustrates how geometrically thin the disc is. The scale of
the radius (x-axis) would have to be multiplied with a factor of
$2.9\times10^{10}$ cm $\Rl^{-1}$ to be translated into the same units
as the scale height.
\label{ophh}}
\end{figure}

We can estimate an equilibrium scale height of the disc (or rather its
chromosphere, Fig.~\ref{ophh}). The value of $H/R$ ranges around 0.01,
i.e.\ the opening angle of the disc is about $0\gdot6$ with a scatter
of $0\gdot1$ and the disc has a scale height of only 1/4 white dwarf
radii at the disc radius. The error on the values of H are fairly small
(less than 1\%), as this parameter only depends on the square-root of
the temperature. This validates our assumption of a geometrically thin
disc. Up to a radius of $0.4\Rl$ the disc has a slight concave shape
with the opening angle ranging between $0\gdot4$ and $0\gdot85$.

However, this disc scale height is calculated with the assumption of
an iso-thermal vertical ($z$-)structure of the disc. Realistic
calculations of the disc photosphere lead (for mass accretion rates
estimated in this paper) to disc opening angles of about 3 degrees in
addition to the underlying disc with unknown, but most likely
non-negligible opening angle. In an as high-inclination system as
V2051~Oph this should lead to significant front-back asymmetries in
the disc emission (however, opposite to what we observe, see
Section~\ref{brightregion}).

\subsubsection{The disc viscosity}
\label{alpha}

Using the formula
\begin{equation}
\frac{3}{2} \Omega_K \alpha P H = \int F_\nu d \nu
\end{equation}
where $\Omega_K = \sqrt{G M_W/R}$ is the Keplerian angular velocity,
$P$ the gas pressure (calculated according to the perfect gas law: $T
= \mu m_H P/\rho k$, with $T$ the temperature, $\mu$ the ionization
coefficient, $m_H$ the mass of the hydrogen atom, $\rho$ the density
and $k$ the Boltzmann's constant), $H$ again the disc scale height and
$F_\nu$ the flux at frequency $\nu$, we can determine the viscosity
parameter $\alpha$ (Shakura \& Sunyaev 1973) throughout the disc. Typical
values of $\alpha$ in a quiescent disc should lie around 0.01 (e.g.\
Smak 1984).

The derived values are very high (between 50 and 900) and incompatible
with current accretion disc models. A very similar situation arises in
the dwarf nova HT~Cas (Wood, Horne \& Vennes 1992, Vrielmann 1997,
Vrielmann et al.\ 2002). We see a need to explain
these high values.

We suggest that the structure of the disc is such, that we do not see the
entire disc. For example, in the case that part of the disc is
obscured or outshined by a hot chromosphere, the gas density is higher
than calculated and $\alpha$ would be lower. In Section~\ref{vertical}
we propose a model of the disc, which could -- in principle -- explain
the observations. However, only concrete theoretical calculations of
such a scenario can show if such a model is indeed applicable.

\section{Summary \& Discussion}
\label{discussion}

\subsection{The distance}
We derived a new distance to V2051~Oph of $146\pm20$~pc using the {\em
Physical Parameter Eclipse Mapping} (PPEM) method, assuming that the
true distance would lead to the smoothest reconstruction. Our result
is compatible with those of Watts et al.\ (1986) and Berriman et al.'s
(1986).  This agreement and a few others others (Vrielmann 2001)
support the applicability of PPEM to derive distances even when we use
a very simple spectral model for the disc emission. However, the
method certainly has to be applied to a large number of systems and
compared to independent distance estimates before it can be considered
established.

\subsection{The radial light distribution of the accretion disc}

We applied the PPEM to a quiescent UBVR data set.  The disc appears
nearly optically thick inside a radius of about $0.2\Rl$ and in the
region where the bright spot is expected (up to a radius of about
$0.4\Rl$), while outside these regions the disc is optically
thin or shows an increasing dominance of a temperature inversion in
the disc chromosphere. This agrees with
Tylenda (1981) and Watts et al.\ (1986) in general, although not in
detail.

Tylenda's (1981) calculations show that a disc with a mass accretion
rate of $10^{16}$\gs\ is optically thin in the outer parts at radii
larger than $1.2 \times 10^{10}$cm (for V2051~Oph this corresponds to
0.42$\Rl$).  Since the accretion disc of V2051 Oph has a radius of
about $1.55 \times 10^{10}$cm ($\sim 0.53\Rl$) this transition zone
lies well within the disc radius, however, is incompatible with ours.

Using the equivalent width (EW) of the Balmer emission lines during an
orbit, Watts et al.\ found the radius of the transition between the
dominance of optically thin over thick regions to be at a radius of
$0.16 \pm 0.02$a (= $0.24 \pm 0.03\Rl$). Their value lies inbetween
our radius of the optically thick part of the disc (0.2 and 0.4$\Rl$),
because their method implies axisymmetry of the accretion disc
emission.

The fact that emission lines of V2051~Oph have broad wings that
originate at radii $R < 0.2\Rl$ (if the disc is Keplerian, see also
also Section~\ref{typesystem}) must mean that they originate from a
(chromospheric) layer above the disc (see Section~\ref{vertical}).

\subsection{The bright region in the accretion disc}
\label{brightregion}

Why does the disc show the {\em bright region} between $-30^\circ$ and
$+60^\circ$ (Fig.~\ref{ophteff}) as well as a high temperature ridge
from the white dwarf to the edge of the disc at an azimuth of about
$-10^\circ$?

This bright region is associated with enhanced emission and therefore
not explicable as a back-front assymetry as expected for a disc with a
significant, underestimated (or neglected) opening angle. The latter
would lead to an enhancement in the ``back'' part of the disc,
i.e. at azimuths around $\pm180^\circ$.

The bright spot region appears to be nearly optically thick in
quiescence and slightly preceding the theoretical position in the
direction of motion in the disc. However, one would rather expect the
bright spot region to extend from the impact area into the direction
of motion in the disc. Even though this bright region might be only
temporary and is only derived using observation near the eclipse, we
discuss its possible origin.

We carefully checked if the reconstructed location of the bright spot
region is caused by an errornous ephemeris, but must conclude that
this cannot be the case.  This means, if the feature is real,
there must be a physical reason for this unexpected disc
structure. Since the secondary plays only a minor role in V2051~Oph,
as its contribution is small compared to the disc, it cannot be
responsible for illumination of the disc.

The high temperature ridge may indicate that the disc is warped and
the ridge at azimuth $-10^\circ$ maximally illuminated by the white
dwarf. The azimuths $+170^\circ$ would then not be illuminated by the
white dwarf.  Since the data we used for the PPEM analysis cover only
the phases $\pm 0.15$ we never see the outer regions of the disc
around azimuth angles $\pm180^\circ$ leading to low temperatures in
our reconstructions. This warp would, however, have to be fixed in the
rotating frame to be so prominent in the maps derived from averaged
light curves. But calculations by Schandl \& Meyer (1994) who propose
coronal winds as the cause of the warp, show a precession of the disc,
making the warp model unlikely.

Another possible scenario is a bulge in the disc similar as
the one proposed by Buckley \& Tuohy (1989) at the location of the
impact region. This bulge could illuminate the disc area around
azimuth angle 0. How the high temperature ridge can be formed is
unclear and it would be interesting to analyse multi-colour
observations of other epochs to see if this feature is stable.

Kaitchuck et al.\ (1994) calculated Doppler images of the emission
lines H$\beta$ and He~I~$\lambda$4471. They find that the leading side
of the disc is brighter than the trailing side. The location of their
bright area does not coincide with our {\em bright region}. However,
they investigate the location of the line emission, which should be
minimal in an optically thick area and could therefore correspond to
the bulge itself or the region ahead of the bulge (i.e.\ at even
larger azimuths) as proposed by Hoard et al.\ (1998) for the nova-like
UU~Aqr. Their disc model involves an emitting impact spot with a trail
of optically thin material along the edge of the disc.

The size and location of the {\em bright region} will be correlated
with the mass accretion rate $\Md$, since the latter determines the
disc size. It would therefore be interesting to study V2051~Oph at
various times during high and low states. For example, the variability
in $\Md$ may also be the reason, why Watts et al.\ (1986) could not
detect the $s$-wave (produced by a bright spot), while it was clearly
present in Honeycutt, Kaitchuck \& Schlegel's (1987) data and might
indicate a variability in the bulge.

\subsection{The vertical disc structure}
\label{vertical}

We propose that the disc itself must have a structure as follows, and
similar to the disc structure suggested for HT~Cas (Vrielmann 2001,
Vrielmann et al.~2002): The disc consists of a cool, optically thick
layer with a hot chromosphere on both sides of the disc. This
chromosphere is nearly optically thick at small radii, but optically
thin for most of the disc.  What we see of the disc is mainly the hot
chromosphere and little from the cool layer underneath, therefore the
temperature and in particular the surface density are descriptive only
of the chromosphere. Such a model was first proposed by Wood et
al.~(1992), Menou (2002) has done some more recent theoretical work on
it.

This disc model resolves in the confusion about the optical depth of
the disc and subsequently our derived values of $\alpha$. On the one
hand, the disc instability model (e.g.\ Ludwig et al.\ 1994, Faulkner
et al.\ 1983, Smak 1982) predicts that the surface density values
should lie between 10 and 100 \gcm\ for $\alpha$'s between 0.1 and 1
and even at larger $\Sigma$'s for smaller $\alpha$'s. This means that
the disc must have a massive, optically thick component.

On the other hand we see line emission from the disc during
quiescence, rise and decline of an outburst (e.g.\ Hessman et al.\
1984) which indicates that there must be an optically thin component
during these phases in the outburst cycle. Our derived surface density
values in the disc re-confirms that the optically thin source (or more
correctly, the non-black body source) is associated with the disc.

The sandwich model elegantly combines both the observational findings
(optically thin material) and the theoretical predictions (optically
thick disc). It also explains why we find such high values of
$\alpha$: Since we {\em see} mainly the contribution from the hot
chromosphere, the surface densities describe only the upper layers of
the accretion disc. The true surface densities ($\Sigma$) must be much
larger, of the order 10 to 100 \gcm. A simple estimate confirms that
values around 100 \gcm\ for $\Sigma$ lead to $\alpha$'s in the
expected range of 0.01 to 1. Thus, we have found a simple explanation,
why we derive such high values for $\alpha$ without discarding the
accretion disc models or our PPEM method: Using the surface density of the
chromosphere alone and assuming it to describe the whole accretion
disc leads to high values of the viscosity parameter $\alpha$. 

The proposed chromosphere must extend to such heights above the
orbital plane that part of it is never eclipsed, i.e.\ near the white
dwarf it must reach farther up than about $0.2\Rl$ or $0.085R_\odot$
or about $8R_{wd}$ (see Section~\ref{sec_unecl}). This chromosphere
should also be the site of the observed emission lines which would be
produced by non-radiative energy dissipation, analogous to a stellar
chromosphere (Horne \& Saar 1991).

A more extensive explanation of the vertical disc structure is
given in Vrielmann et al.~(2001) for the similar system HT~Cas.

\subsection{The mass accretion rate}
\label{massacc}

Similar to results of Watts et al.\ (1986), the quiescent mass
accretion rate appears in large parts of the disc to be somewhat
higher than or around the critical values given by Ludwig et al.\ 1994
or Watts et al.\ of $\sim 4.6 \times 10^{-11} \Msol$~yr$^{-1}$ (for
$K_1 = 111$km s$^-1$). The latter derived a similar accretion rate for
V2051~Oph as we of $\sim 1.3 \times 10^{-10} \Msol$~yr$^{-1}$ compared
to our $1.6 \times 10^{-11}\Msol$~yr$^{-1}$ to $1.6 \times
10^{-9}\Msol$~yr$^{-1}$.  Fig.~\ref{ophteff} shows that specifically
the bright spot and inner, hot regions ($R<0.1\Rl$) have higher
temperatures than the critical values $\TeffA(R)$.

Since V2051~Oph shows outbursts, and probably very infrequently, the
effective temperature should not lie so close to or even above the
critical $\TeffA(R)$, but rather far below (e.g. Warner 1995).
Berriman et al. (1986) also found that V2051~Oph's disc is too hot to
undergo the disc instability cycle. The only known outbursts around
the time of our observations (June 1983) were the ones in April 1982
and in May 1984. V2051~Oph has infrequent observed outbursts and a
typical outburst cycle of about a few years (VSNET
observations). So the disc is unlikely to have been affected by a rise
to or decline from outburst. A similar situation occurs in HT~Cas
(Vrielmann et al.\ 2002), a dwarf nova with reportedly infrequent
outbursts. At the moment, we have no solution for this problem.
Until another explanation is found, we must assume that either our
mass accretion rates are errorneous or that it is sufficient if part
of the disc has effective temperatures below the critical value.

We would recommend a study of this object throughout a quiescent cycle
in order to check variations in the mass accretion rate, especially
before the onset of an outburst and during low states ($B\sim 16.2$,
Baptista et al.\ (1998), compared to $B\sim 15$ (WO). Interestingly,
HT~Cas, a similar dwarf nova, also shows variations during quiescence
(Robertson \& Honeycutt 1996) with fluctuations of about 1.5 mag.

Furthermore, it would be very desirable to get an independent distance
estimate. If the distance were significantly smaller than our
estimate, the effective temperatures and mass accretion rates would be
reconstructed to smaller values. However, this will not solve the
problem, because an (unlikely) reduction of a factor two in distance
will only lead to a small reduction in effective temperatures to
$0.5^{1/4}\Teff = 0.84 \Teff$ and therefore to a small change in
$\Md$.

\subsection{Is V2051 Oph an intermediate polar?}
\label{typesystem}

While it is clear that V2051~Oph shows typical features of a SU~UMa
system (outbursts and super outbursts), it is not clear, whether the
white dwarf has a small magnetic field.

An intermediate polar should show a disrupted inner disc in the PPEM
maps, unless the inner disc radius is too small. We do not see any
indication for a hole, except for a flattening of the effective
temperature towards the white dwarf. Only a very tiny hole could be
hidden due to the MEM smearing, as tests of the PPEM method
show. However, estimating the line width at phase 0.51 of H$\alpha$ in
WO's Fig.~10, we yield $\sim77$\AA, i.e.\ the disc material at the
inner disc edge has a velocity of about 1777 \kms. For a Keplerian
disc this corresponds to an inner disc radius of $3.3\times10^9$ cm $=
0.11 \Rl$. More recent, yet unpublished data of V2051~Oph suggest line wings
extending to about 2530 \kms, corresponding to an inner disc radius of
2.3 white dwarf radii. Quiescence data of Watts et al.~(1986) also
show line wings of H$\beta$ and H$\gamma$ extending to maximally
2500~\kms\ (corresponding to $R_{in} = 0.06\Rl = 2.4\Rwd$).

Steeghs et al.~(2001) find much broader line wings. Both spectra sets
were taken during decline from outburst which means that the emission
lines should have been affected in the same way. Since the intensity
in the outer line wings is extremely low ($<3\%$ for $v < -2500$~\kms;
red wing affected by He line, Fig. 3 of Steeghs et al.), broad shallow
wings could -- in principle -- be lost in the noise of the above
mentioned spectra, however, no such line profiles as observed by
Steeghs et al.\ can be hidden in those spectra. Only the blue wing of
H$\beta$ can be measured reliably, which will be enhanced by the disc
wind, a possible explanation for Steeghs et al.'s broad line profile.

The fact that we do not see a hole in the reconstructed disc means
either, there is none, it is extremely small, or it could be hidden by
accretion curtains.

X-ray observations (Holcomb, Caillault \& Patterson 1994, van
Teeseling, Beuermann \& Verbunt 1996) and lack of circular
polarisation (Crop\-per 1986) data do not contradict the intermediate
polar model with a very weak magnetic field of 1~MG or less, as
expected for intermediate polars.

Thus, if V2051~Oph is indeed an intermediate polar, its white dwarf
must have a very weak magnetic field which allows to disc to be
disrupted with an inner disc radius of $R_{in} < 0.1\Rl = 4.2\Rwd$. At
present we cannot distinguish between such a low field intermediate
polar and a dwarf nova. However, the variable hump and irregular
occurance of strong flares as observed by WO and Berriman et al.\ 1986
might indicate a flaring accretion curtain which lights up during
inhomogeneous accretion. Warner \& Woudt (2002) propose a model for
V2051~Oph in which the spinup of the magnetic field lines of a small
field on the white dwarf temporarily creates such a small hole of a a
few tenths larger than the white dwarf radius. Their model applies to
a disc on decline from outburst. During quiescence they expect the
hole to be even slightly larger. If this hole is present and
persisting long enough, accretion curtains can develop, leading to the
observed flares.

Simultaneous multi-colour observations at different epochs and high
resolution spectroscopy are very desirable in order to solve the
mystery of V2051~Oph's nature.

\subsection{Comparison to similar objects}
\label{comparison}

V2051~Oph has a similar orbital period as OY~Car, Z~Cha and HT~Cas and
since all of them are SU~UMa system, showing super outbursts inbetween
their normal ones, one would expect to find some similarities among
these systems. However, the light curve of V2051~Oph seems to vary a
lot more than that of the other three systems, especially in terms of 
the presence and location of the (orbital) hump, flickering and flaring. 

The {\em eclipse} light curve of OY~Car (Vogt et al.\ 1981, Wood et
al.\ 1989, Rutten et al.\ 1992, Horne et al.\ 1994) is remarkably
stable over more than a decade and even close to a super outburst
(Hessman et al.\ 1992). A similar consistency in the appearance is
shown by HT~Cas (Berriman et al.\ 1987, Horne et al.\ 1991, Robertson
\& Honeycutt 1996) and Z~Cha (Wood et al.\ 1986, Marsh et al.\ 1987,
Robinson et al.\ 1995). OY~Car and Z~Cha even show text book light
curves of eclipsing dwarf novae in which the orbital hump and the
ingress and egress phases of the white dwarf and the bright spot can
be very well distinguished and explained in the standard Roche model.
HT~Cas, on the contrary, has never been observed to show an orbital
hump.

\begin{table*}
\caption{System parameters of V2051~Oph and similar SU~UMa type dwarf
novae in comparison.
\label{tab_para}}
\vspace{1ex}
\begin{tabular}{lcccc}
Parameter \hspace{2cm} & \hspace{3ex}V2051~Oph\hspace{3ex} & \hspace{7ex}HT~Cas\hspace{7ex} & \hspace{7ex}OY~Car\hspace{7ex} & \hspace{7ex}Z~Cha\hspace{7ex} \\ \hline
$P\hspace{1ex} [h]$                &  1.50     & 1.77   & 1.51   & 1.79 \\

$i$ & $83\gdot3\hspace{1ex}\pm\hspace{1ex}1\gdot4$ & $81\gdot0\hspace{1ex}\pm\hspace{1ex}1\gdot0$& $83\gdot3\hspace{1ex}\pm\hspace{1ex}0\gdot2$ & $81\gdot7\hspace{1ex}\pm\hspace{1ex}0\gdot13$\\

$q$             & $0.19\hspace{1ex}\pm\hspace{1ex}0.03$  & $0.15\hspace{1ex}\pm\hspace{1ex}0.03$ & $0.102\hspace{1ex}\pm\hspace{1ex}0.003$ & $0.1495\hspace{1ex}\pm\hspace{1ex}0.0035$ \\

${\cal M}_{wd}\hspace{1ex} [\Msol]$ & $0.78\hspace{1ex}
\pm\hspace{1ex} 0.06$  & $0.61\hspace{1ex}\pm\hspace{1ex}0.04$ &
$0.685\hspace{1ex}\pm\hspace{1ex}0.011$ &
$0.54\hspace{1ex}\pm\hspace{1ex}0.01$ \\

${\cal M}_r\hspace{1ex} [\Msol]$    & $0.15\hspace{1ex}\pm\hspace{1ex}0.03$  & $0.09\hspace{1ex}\pm\hspace{1ex}0.02$ & $0.070\hspace{1ex}\pm\hspace{1ex}0.002$ & $0.081\hspace{1ex}\pm\hspace{1ex}0.003$ \\

$\Md$\hspace{1ex} [gs$^{-1}$] & $1\times10^{17}$ & $5\times10^{15}$ & $4\times10^{15}$ & $4\times10^{15}$\\
\hline
\end{tabular}
{
\begin{tabbing}
$\Md$: Mass accretion rate at the disc radius (disc edge) or into the bright spot\\
References: {\bf V2051~Oph:} Baptista et al.\ (1998), {\bf HT~Cas:} Horne et
 al.\ (1991), Vrielmann et al.\ (2002, $\Md$),\\
{\bf OY~Car:} Wood et al.\ (1989),
Horne et al.\ (1994), {\bf Z~Cha:} Wood et al.\ (1986)\\
\end{tabbing}
}
\end{table*}

For comparison, we listed a few system parameters of these four
objects in Tab.~\ref{tab_para}.  While at first glance all parameters
seem to be similar, closer inspection reveals that the secondary in
V2051~Oph is about twice as massive as the secondaries in the other
systems. Furthermore, also the mass accretion rate $\Md$ into the disc
of V2051~Oph is larger by a factor of twenty or more than in the other
systems (in OY~Car and Z~Cha $\Md$ was determined by Eclipse Mapping
(Wood et al.\ 1989 using d=100~pc, Horne et al.\ 1994; Wood et al.\
1986) and in HT~Cas with the PPEM method (Vrielmann et al.\ 2002)).

The larger amount of variation of V2051~Oph could be induced by the
larger mass accretion rate, being close to or above the theoretical
limit. Even if the disc were rather patchy with a covering factor of
maybe 50\%, as proposed for HT~Cas (Vrielmann et al.\ 2002), the mass
accretion rate were still a factor of about ten or more larger than
that of the other three dwarf novae. The larger mass of the secondary
in V2051~Oph, {\em could} be the reason for the higher mass accretion
rate, however, it is not fully understood, how $\Md$ is related to
other system parameters. Furthermore, it is not clear, how the larger
$\Md$ alone can cause the occational occurence of a hump. We
therefore, cannot exclude the low-field intermediate polar model.

\subsection{Concluding remark}

Since V2051~Oph is usually too faint for amateur astronomers,
monitoring of this object is difficult, but particularly interesting.
Especially also the evolution of the disc during normal and low
quiescent states and the percentage of time V2051~Oph spends in either
state would be of immense interest.

\section*{acknowledgments}
We are indebted to Keith Horne for inspecting, calibrating, reducing
and communicating the data. Furthermore, we thank Brian Warner,
Stephen Potter, Encarni Romero Colmenero, K.\ Horne and Matthias
Schreiber for insightful discussions and B.\ Warner for corrections of
an intermediate draft. Finally, many thanks go to the anonymous
referees for comments leading to a substantial improvement of this
paper. This work is funded by the South African NRF and the CHL
foundation through postdoctoral fellowships for SV.

\end{document}